# Brief introduction to discrete Boltzmann modeling and analysis method

Aiguo Xu

National Key Laboratory of Computational Physics, Institute of Applied Physics and Computational Mathematics, P. O. Box 8009-26, Beijing 100088, P.R.China

HEDPS, Center for Applied Physics and Technology, and College of Engineering, Peking University, Beijing 100871, P.R.China

## Abstract

We briefly introduce several fundamental problems that cause the creation of Discrete Boltzmann modeling and analysis Method(DBM), corresponding solutions, the relationship and difference between DBM and traditional fluid modeling and other kinetic methods, and some applications of DBM and discrete/non-equilibrium effects.

## 1. Academic background

Due to the heterogeneity of the medium and the diversity of the environment, what the shock and detonation bring are often the typical multi-scale non-equilibrium complex flows. Such systems often contain spatial structures and kinetic patterns of different scales. The existence of these intermediate-scale structures and patterns significantly affects the physical properties and functions of the system. The large scale, slow change behavior can generally be well described by Navier-Stokes (NS) equations. However, in the description of flow behavior in some low-pressure sparse regions, the description of internal structure of shock wave or detonation wave, and the description of non-equilibrium behavior caused by fast changing flow or reaction, NS equations are insufficient in physical function. At the same time, microscopic Molecular Dynamics (MD) simulations are often powerless to the flow behavior we care about due to the limitations of applicable spatiotemporal scales. Therefore, due to the lack of appropriate research methods, the cognition of these mesoscale behaviors is extremely weak, which greatly affects the evaluation of various effects of "mesoscale" behaviors, the development of physical functions and the development of corresponding regulatory technologies. The modeling and mechanism analysis of mesoscale behavior are the core contents of current *mesoscience* research [1].

## 2. Problems and challenges

The discontinuity or discreteness of a fluid system is closely related to the degree of

Thermodynamic Non-Equilibrium. The Knudsen number (Kn), on the one hand, can be regarded as the average molecular distance of re-scaling, describing the discreteness or discontinuity of the system. On the other hand, it can also be regarded as the thermodynamic relaxation time of re-scaling to describe the degree of TNE of the system. The concepts of non-equilibrium flow and discontinuous flow (discrete flow) overlap in physical connotation. The study of non-equilibrium complex flows has made great progress, but still faces some fundamental scientific challenges, such as:

(i) Cross-scale modeling and simulation, which has been a research hotspot for about 15-20 years. During this period, there have also been ups and downs in people's enthusiasm. The main methodology is to modify the macroscopic fluid/solid equations. That is, to grasp the state and behavior of the system, the same physical quantities used in traditional fluid/solid mechanics equations. A natural basic science question, then, is: as the degree of discreteness/non-equilibrium increases, is it really okay to focus only on the few physical variables in a continuous model (e.g. NS)?

(ii) It is a typical feature that complex structures and behaviors generally take on different characteristics when viewed from different angles. The methods for how to extract more valuable information and how to analyze determines our research ability and depth. Among them, the technical key is as below: in the face of the increasing complexity of the system, how to achieve the intuitive geometric correspondence of the complex system state and behavior description?

(iii) The turbulence research, because of its importance and become the unwavering research hotspot. Turbulent mixing is an important part of hydrodynamic instability research. It can be thought that as the vortex and other structures concerned become smaller and smaller, the average molecular distance is no longer a negligible small quantity relative to the scale of the structure or behavior concerned, that is, the discretization will become stronger and the discretization effect will become more and more significant. However, the early concepts and theoretical framework of turbulence are based on the continuum theory. So, the third fundamental scientific question is: Is the concept of turbulence based on the macroscopic continuous image mathematically and the concept of turbulence based on the pursuit of physical origin exactly the same?

(iv) The small system heat and mass transfer appears to have some abnormal behavior, being different from those in the macroscopic continuous situation, has been attracting more attention and research in the field of statistical physics and nonlinear science. However, the small systems widely studied in the literature of statistical physics and nonlinear science are often too small compared to the mesoscale non-equilibrium cases concerned by the mechanical engineering community. We are faced with a situation where engineering applications adopt mostly macroscopic continuous modeling, but also find some unsatisfactory or even unsolvable problems. The question we face is: how can mesoscale modeling be seamless with macroscopic continuous modeling?

(v) In the past years, the main idea of cross-scale modeling and simulation is reductionism. Reductionism naturally brings a lot of help, brings a lot of physical cognition. It is true that along the reductionist train of thought, our research has made great progress. But it is a characteristic of complex systems that, as complexity increases, new behavioral features emerge/emerge/evolve that cannot be deduced from simpler or simplest cases by any composition or coexistence laws. (Emergence, the idea that the whole is greater than the sum of its parts, is a central theme in complexity science.) So, the fifth fundamental scientific question that DBM needs to face is: Is reductionist thinking really enough for cross-scale modeling and simulation?

(vi) The sixth basic science problem relates to the most basic parameter of non-equilibrium flow description/cognition, that is the non-equilibrium strength. Given the complexity of non-equilibrium behavior, any definition of non-equilibrium strength depends on the research perspective. In other words, there may be the following situations: the non-equilibrium strength is increasing from a certain perspective, while from another perspective, it's going down. How to describe the non-equilibrium strength of complex flows?

## 3. Ideas and schemes

Simulation study of complex flow includes three major steps, as shown in Fig. 1, (i) physical modeling, (ii) discrete format selection/design, (iii) numerical experiment and complex physical field analysis. As a user of discrete format, the focus of physical mechanics research groups is generally on steps (i) and (iii). The best discrete format design in step (ii) is left to computational mathematicians. Besides performing numerical experiments according to the research need, the task of physical mechanics research groups is two-fold: (i) to ensure the rationality of the physical model (theoretical model) for the problem to be studied, taking into account the simplicity; try to extract more valuable physical information for the complex physical field from the massive data.

### 3.1 Brief introduction of DBM

Statistical physics is a bridge between micro and macro, and the kinetic theory is a bridge between micro and macro descriptions of fluid systems. Coarse-grained modeling is a basic means to grasp the main contradiction according to the research needs of non-equilibrium statistical physics. The theory on Boltzmann equation is a relatively mature part in the theory of kinetics. The term "cross-scale" in kinetic description generally refers to the cross-Knudsen number.

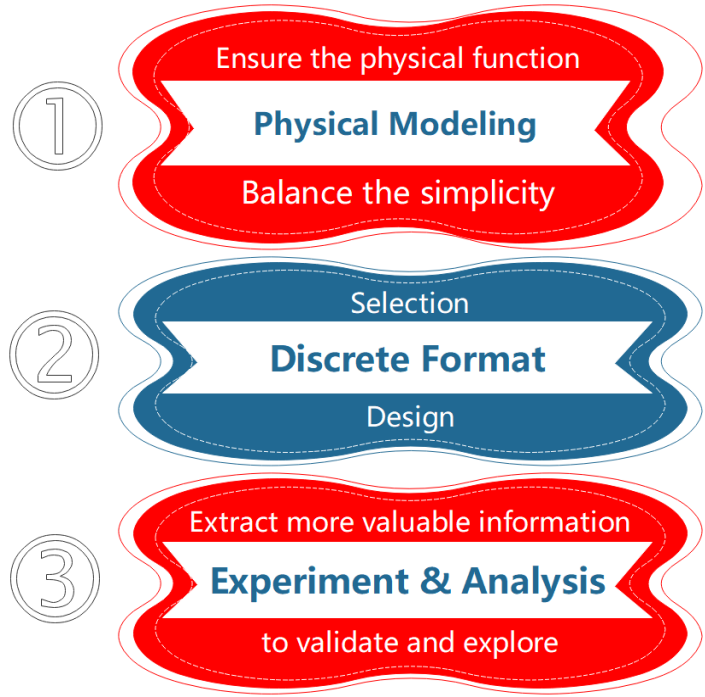

Fig. 1: Three major steps of simulation study of complex flows.

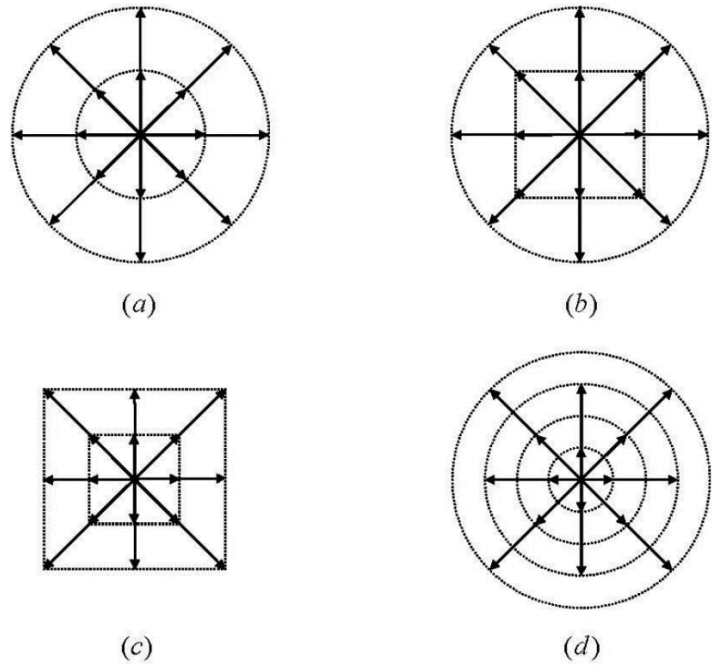

Fig.2: Schematic of several sets of commonly used discrete velocities.

The Discrete Boltzmann Method (DBM) is a physical model construction and complex field analysis method based on the discrete Boltzmann equation,

$$\frac{\partial f_i}{\partial t} + \mathbf{v}_i \cdot \frac{\partial f_i}{\partial \mathbf{x}} + (\text{force term})_i = (\text{collision term})_i \qquad (1)$$

and aiming to (partly) attack the above several basic scientific problems. Where the subscript "*i*"of the distribution function $f$ is the index of the discrete velocity, corresponding to the discrete velocity $\mathbf{v}_i$. Obviously, *the rule of discrete velocity selection is the key technology in DBM modeling.* Figure 2 shows the schematic diagram for several sets commonly used discrete velocities.

*In terms of physical model construction, DBM includes two steps of coarse-grained physical modeling: (i) modification and simplification of Boltzmann equation and (ii) discretization of particle velocity space.* The basic principle of coarse-grained physical modeling is that the behavior of the system to be studied cannot be changed by the simplification of the model. In the theory of kinetics, in addition to the distribution function $f$ itself, the properties of the system are described by the kinetic moments of the distribution function. Therefore, the kinetic moments corresponding to the kinetic properties to be studied given by the governing equations must be the same before and after the simplification of the collision term and the discretization of the particle velocity space. After the collision term is simplified, DBM gives the most necessary physical constraints for the selection of discrete velocity: the kinetic moments involved in the description of the system behavior are converted from integral $\int f^{(0)} \mathbf{\Psi}(\mathbf{v}) d\mathbf{v}$ to summation $\sum_i f_i^{(0)} \mathbf{\Psi}(\mathbf{v}_i)$ for calculation, and the results remain unchanged, that is

$$\int f^{(0)} \mathbf{\Psi}(\mathbf{v}) d\mathbf{v} = \sum_i f_i^{(0)} \mathbf{\Psi}(\mathbf{v}_i). \qquad (2)$$

But it does not adhere to the specific discrete format. Here, $f^{(0)} = f^{eq}$, is the corresponding equilibrium distribution function.

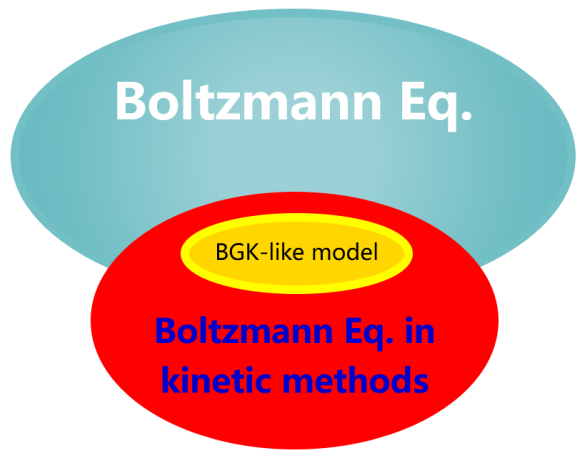

Fig.3: Schematic of original Boltzmann equation and the Boltzmann equation in kinetic methods.

It should be pointed out that many behaviors of complex flows are beyond the description capacity of the original Boltzmann equation. The Boltzmann equation in DBM is actually the modified Boltzmann equation combined with the mean field theory according to the specific situation, and its application scope may be larger than the original Boltzmann equation in some aspects, as shown in Figure 3 [2-3]. Therefore, DBM is actually a coarse-grained modeling and analysis method that combines kinetic theory and mean field theory. When dealing with the external force term of the Boltzmann equation, in order to avoid the difficulty of differentiating the discrete velocity, DBM modeling is: first approximate the external force term, by using property that the equilibrium distribution function $f^{(0)}$ is derivable to the particle velocity $\mathbf{v}$, complete the derivative of the particle velocity, and then write in the form of discrete velocities.

*In terms of complex physical field analysis methods, DBM is the specific application and further development of statistical physics coarse-grained description method, non-equilibrium behavior description method, and phase space description method under the framework of discrete Boltzmann equation.* Historically, DBM has evolved from the physical modeling branch of the Lattice Boltzmann Method (LBM) with some additions and some abandons. Keep the use of discrete velocity, but no longer adhere to the specific discrete format, only give the most necessary physical constraints to the discrete format. The DBM no longer bases on the continuum assumption and near equilibrium hypotheses, no longer uses the "lattice gas" physical image of the standard LBM, adds methods for detection, presentation, description and analysis of non-equilibrium states and resulting effects based on phase space, and more information extraction techniques and complex physical field analysis techniques will be introduced with time. Figure 4 shows the development from the phase space description method based on $(f - f^{(0)})$ non-conservative moments,

$$\begin{aligned}
\mathbf{\Delta}_n^* &= \mathbf{M}_n^* \left( f - f^{(0)} \right) \\
&= \int d\mathbf{v} \left( f - f^{(0)} \right) \underbrace{(\mathbf{v} - \mathbf{u})(\mathbf{v} - \mathbf{u}) \cdots (\mathbf{v} - \mathbf{u})}_{n\text{-th order tensor}} \\
&= \sum_i \left( f_i - f_i^{(0)} \right) \underbrace{(\mathbf{v}_i - \mathbf{u})(\mathbf{v}_i - \mathbf{u}) \cdots (\mathbf{v}_i - \mathbf{u})}_{n\text{-th order tensor}}
\end{aligned} \quad (3)$$

to the phase space description method based on any set of characteristics,

$$\mathbf{X} = \{X_1, X_2, X_3, \cdots\}. \quad (4)$$

In Eq.(3), $\mathbf{u}$ is the local flow velocity. In the phase space or its subspace, we can define the non-equilibrium intensity (or the intensity of the corresponding behavior feature) of the corresponding viewing angle with the help of the distance $D$ from the state point to the coordinate origin, as shown in Figure 4. Use the concept of distance between two points to describe the difference between two non-equilibrium states (or the corresponding behavioral characteristics), such as $d$ in Figure 4, and so on. Therefore, a more complete name for DBM is Discrete Boltzmann

*modeling and analysis Method [2-5]. The use of non-conservative kinetic moments to detect and describe the specific ways in which a system deviates from thermodynamic equilibrium and the various effects caused by this is the key technology of DBM in the analysis of complex physical fields.*

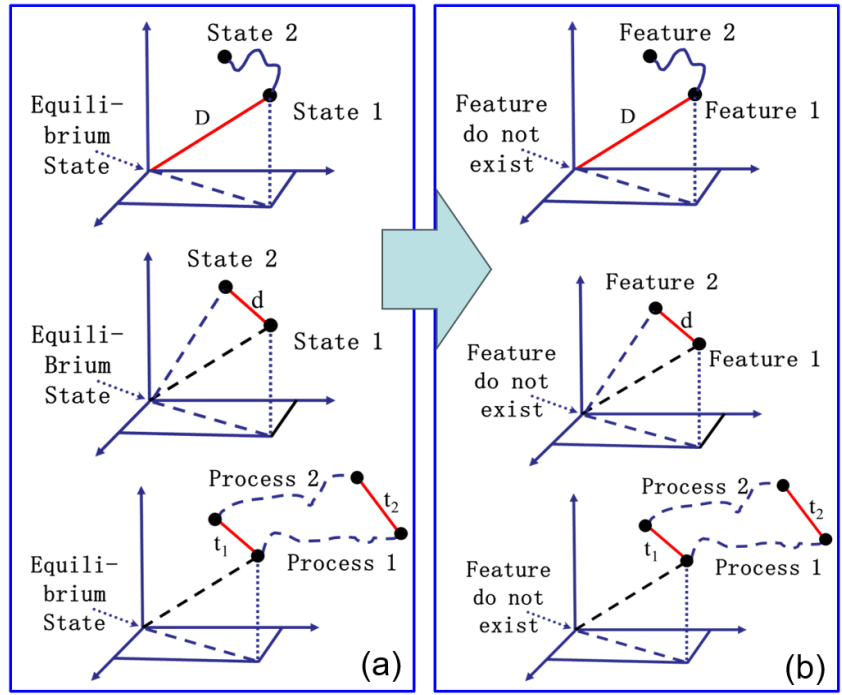

Fig.4: Development of phase space description method from that based on $(f - f^{(0)})$ non-conservative moments (a) to that based on any set of characteristic features (b).

The behavior of real systems is often complex. Coarse-grained modeling is a process of "losing information", but this "losing information" has a bottom line. The bottom line is that the nature of what needs to be studied cannot be changed by simplifying the model. DBM, according to the research needs of the problem, choose a perspective to study a set of kinetic properties of the system, and require the kinetic moments describing this set of properties to preserve value in the process of model simplification. With the increase of the degree of discreteness and non-equilibrium, the complexity of system behavior rises sharply, and more high-order kinetic moments enter the description is the inevitable result of the requirement that the control ability does not decline, so the use of more physical quantities to describe the system state and behavior is a typical feature of DBM that is different from traditional fluid modeling and other current kinetic methods. From the perspective of Kinetic Macro Modeling (KMM), this is the requirement to obtain more accurate constitutive relations. From the perspective of kinetic theory, this is the requirement to obtain a more accurate distribution function. Different perspectives lead to the same destination.

The starting point for DBM is the side of the "mesoscale" range near the macro. Due to the stage of development, the main cases considered by DBM at present are those where Chapman-Enskog multi-scale analysis theory is valid. Therefore, Chapman-Enskog multi-scale analysis theory is a reasonable and effective mathematical guarantee for DBM thinking [2].

**3.2 Rule of selecting discrete velocities**

To calculate the conservative kinetic moments, i.e., the density, momentum and energy moments, on both sides of the Boltzmann equation at the same time, we can obtain a set of hydrodynamic equations consistent with the form of NS, which we call generalized NS. The *advantage of this generalized NS* is that it does not approximate its stress

$$\mathbf{\Delta}_2^* = \mathbf{M}_2^* \left( f - f^{(0)} \right) = \mathbf{M}_2^* \left( f^{(1)} + f^{(2)} + f^{(3)} + \cdots \right) \tag{5}$$

and heat flux

$$\mathbf{\Delta}_{3,1}^* = \mathbf{M}_{3,1}^* \left( f - f^{(0)} \right) = \mathbf{M}_{3,1}^* \left( f^{(1)} + f^{(2)} + f^{(3)} + \cdots \right) \tag{6}$$

where all orders of the non-equilibrium distribution functions, $f^{(j)}$, are included, where $j = 1, 2, 3, \cdots$ corresponds to the order of discreteness/TNE, i.e., the power of Knudsen number. The meaning of the subscript "3,1" is that is the 1st-order tensor contracted from the 3th-order tensor, that is, a vector, after one dot multiplication. The meaning of the rest of the similar subscripts below, and so on. *The disadvantage* (of this generalized NS) is that the specific expressions of stress and heat flow cannot be written, so there is no way to use them directly. To overcome the inability to write constitutive relations, Chapman and Enskog developed the Chapman-Enskog multi-scale analysis method, which was later named after them. Chapman-Enskog multi-scale analysis is actually a generalized Taylor expansion and analysis method: the independent variable here is Kn number. Not only the distribution function, but also the time and space derivatives, do Taylor expansions at point, $\text{Kn} = 0$ (i.e., at continuous, thermodynamic equilibrium state). More physical image interpretations of Chapman-Enskog multi-scale analysis can be found in reference [2]. Chapman-Enskog multi-scale analysis tells us:

(i) The hydrodynamic equations that the Boltzmann equation corresponds to in the limit of quasi-continuum and under near equilibrium condition are the NS. The viscous stress and heat flux that consider only the contribution of the 1st order nonequilibrium, $f^{(1)}$, are the NS viscous stress and heat flux. With increasing the Kn number, i.e., with increasing the degree of discreteness/TNE, the more accurate viscous stress and heat flux should include the contribution of the 2nd order $f^{(2)}$, and even higher order nonequilibriums.

(ii) Of all the kinetic moments of the distribution function $f$, entering NS describes three

conserved moments (density $\rho$, momentum $\rho\mathbf{u}$, and energy $E$) and two non-conserved moments (viscous stress and heat flux). The remaining non-conserved moments do not enter the NS description, which is a double-edged sword: on the one hand, it brings the simplicity of traditional fluid mechanics theory. However, on the other hand, it sets up a barrier for NS to describe the more discrete/nonequilibrium case, and is the physical reason why NS does not describe the more discrete/nonequilibrium case.

(iii) The ability to recover the corresponding level of macroscopic fluid mechanics equations (such as NS, Burnett equations) is only part of the physical function of DBM. Corresponding to the physical function of DBM is the Extended Hydrodynamic Equations (EHE), that is, in addition to the conserved moment evolution equations corresponding to the three conservation laws of mass, momentum and energy, the system also includes some of the most closely related non-conserved moment evolution equations. The modeling method of EHE derivation based on kinetic equations is called Kinetic Macro Modeling (KMM). The necessity of the extended part, i.e. the evolution equations of the relevant non-conserved moments, rapidly increases as the degree of discreteness/non-equilibrium increases.

(iv) As the degree of discreteness/non-equilibrium represented by the Knudsen number increases, the complexity of DBM simulations increases relatively slowly compared to KMM simulations, i.e. deriving and solving EHE, and can therefore go further. If the order of Knudsen number to be considered increases by 1, the number of kinetic moments that need to be preserved in DBM increases by 2. However, the complexity of KMM's derivation of EHE generally increases sharply. In the case of multi-components, the forms of EHE are not unique because of the respective flow velocities and average flow velocity of different components, and the definition of temperature depends on the reference flow velocity. These forms of EHE correspond to the descriptions of complex multi-component flows from different perspectives. It can be seen that in the multi-component case, the corresponding relationship between DBM and physically functionally equivalent EHE is one-to-several. Here, several means more than one.

(v) DBM simulation, it is not necessary to know the specific form of the extremely complex EHE of physical functional equivalence, but we can borrow the ideas and physical images of KMM, without strictly deriving the equation, just look at the order of the kinetic moments, to quickly screen out the kinetic moments involved in more accurate constitutive relations, i.e., viscous stress and heat flux calculations. This set of the kinetic moments are converted from integral to sum for calculation, and the results need to remain unchanged. At this point, we obtain the most necessary physical constraint for the discrete velocity selection: the system of linear equations represented by equation (2).

### 3.3 Description of nonequilibrium behavior

Non-equilibrium strength is one of the most basic parameters in the description and cognition of non-equilibrium flow. In addition to the commonly used Knudsen number, spatial gradients and time rates of change of macroscopic physical quantities such as the density, flow velocity, temperature, pressure, etc., DBM uses non-conserved kinetic moments of $(f - f^{(0)})$ to describe the way and amplitude of deviation from equilibrium, and further uses non-conserved moments $\{\Delta_n^*, n = 2, (3,1), 3, (4,2), 4, (5,3), \cdots\}$ (independent components) of $(f - f^{(0)})$ as the basis to open phase space. Provide an intuitive geometric mapping of complex system state and behavior characteristics (as shown in Figure 4). It is easy to imagine that each of the non-conserved moments of $(f - f^{(0)})$ itself or any of its independent components describes the manner and magnitude of the system's deviation from equilibrium from its own perspective, thus providing a perspective of the non-equilibrium strength. The non-equilibrium intensity of these different perspectives, related, complementary, and generally unsubstitutable, together form a more complete description. It should be emphasized that as a later, more basic description method, DBM naturally inherits all the traditional methods of describing nonequilibrium behavior.

Any definition of non-equilibrium strength depends on the research perspective. The non-equilibrium strength is increasing from a certain perspective, while from another perspective, it may be decreasing. This is one of the concrete manifestations of the complexity of nonequilibrium behavior. If we only study the nonequilibrium strength and the resulting effects of one perspective, the understanding obtained is often one-sided or even wrong. (Being wrong refers to the misconception that the conclusions drawn are not dependent on the research perspective and are universal.) In view of this, based on non-conserved moments of $(f - f^{(0)})$, macroscopic quantity gradients, thermodynamic relaxation time, Knudsen number, morphological description, etc., DBM introduces a multi-view non-equilibrium description scheme to carry out a multi-view cross-positioning of non-equilibrium strength of complex flows. For the convenience of description, the concept of non-equilibrium intensity vector is further introduced, where any component of the vector is a non-equilibrium intensity of a corresponding viewing angle. For example, the composition of a non-equilibrium strength vector might be

$$\mathbf{S} = \{D, D_2, \nabla \rho, \nabla T, \tau, \text{Kn}, \cdots\} \tag{7}$$

where

$$D_2 = \sqrt{\Delta_{2,xx}^{*2} + \Delta_{2,xy}^{*2} + \Delta_{2,yy}^{*2}} \tag{8}$$

is the distance in the corresponding subspace of the phase space based on the independent components of $\left(f-f^{(0)}\right)$ non-conservative kinetic moments. It is a nonequilibrium strength contributed by the perspective of $\mathbf{\Delta}_2^*$. $T$ is the local temperature, and $\tau$ is the thermodynamic relaxation time.

Just as the phase space description does not require its coordinate axes to be equidimensional, the non-equilibrium intensity vector does not require its components to be equidimensional. As a set of behavior characteristics, the non-equilibrium strength of different viewing angles can also be described by the phase space method, which provides an intuitive geometric correspondence for the non-equilibrium strength of different viewing angles in complex flows. Of course, as a coarse-grained modeling and description method, the physical accuracy of DBM, including the description of non-equilibrium strength, needs to be adjusted according to specific research requirements [2].

### 3.4 A few notes

At a Report on the 2nd U.S.-Japan Joint Seminar, held at Santa Barbara, California, from 7-10 August, 1996, late eminent scholar, Mr. Chang-Lin Tien, et al. commented the LBM as follows: "Many physical phenomena and engineering problems may have their origins at molecular scales, although they need to interface with the macroscopic or 'human scales'. The difficulty arises in bridging the results of these models across the span of length and time scales. The lattice Boltzmann method attempts to bridge this gap." More details are referred to Ref. [6]. Many people's understanding of LBM is directly or indirectly affected by this or similar comments.

In fact, starting from its predecessor lattice gas (cellular automata) method [7-8], the LBMs in the literature have two categories: coarse-grained physical model construction method and numerical solving method to some governing equations[9]. The latter accounts for the vast majority in the existing literature, so much so that LBM has almost become a shorthand or pronoun for the latter[10-12]. However, the LBM mentioned by Mr. Chang-Lin Tien, et al. in the above comment is obviously a cross-scale physical model construction method, which is the former. Coarse-grained physics modeling is the basic means of non-equilibrium statistical physics to grasp the main contradiction according to the research needs, and the physical function inherited from the Boltzmann equation description across scales. In this case, across Kn numbers. These two different types of LBM, working in complementary dimensions, different goals, different construction rules, each reasonable.

In the absence of ambiguity, DBM is often used separately as short for Discrete Boltzmann Method, Discrete Boltzmann Modeling, and Discrete Boltzmann Model. As mentioned earlier, the Discrete Boltzmann Method is more fully called the Discrete Boltzmann modeling and analysis Method. When compared with other methods, the identity of DBM is "physical model construction method + complex physical field analysis method", or the physical model construction method with complex physical field analysis function. When compared with other physical models, DBM is a physical model with complex physical field analysis capabilities. At present, the commonly referred physical model does not provide complex physical field analysis function for simulated data, so when the name discrete Boltzmann model or modeling is mentioned, people usually do not realize that it also has complex physical field function, which is the reason why our research group has to call it physical model construction and complex physical field analysis method in some occasions.

*The physical model building part of DBM includes determining the specific governing equations and giving the most necessary physical constraints to follow for the discrete velocity selection.* Just like NS simulation, in DBM simulation, the specific discrete formats of spatial derivatives, time integrals and particle velocities need to be reasonably selected according to specific conditions and flow patterns.

According to equation (3), from the perspective of dimensions and physical images, the physical meaning of non-equilibrium feature quantity is Non-Organized Flux. It should be noted that in a macroscopic description, when a physical quantity (such as density $\rho$) is multiplied by the flow velocity $\mathbf{u}$, its physical meaning is the flux of this physical quantity, which is an organized flux. *The organized and non-organized fluxes together form a more complete description of the kinetic behavior of complex flows.* Of all the non-organized fluxes composed of non-conserved central moments of $(f - f^{(0)})$, only the Non-Organized Momentum Flux (NOMF) $\mathbf{\Delta}^*_2$ and the Non-Organized Energy Flux (NOEF) $\mathbf{\Delta}^*_{3,1}$ correspond in NS. Corresponding to viscous stress and heat flux, respectively. The rest of the non-conserved central moments have no direct correspondence in NS. But, as mentioned earlier, their physical meaning is clear. The absence of these physical quantities in the NS description is precisely the physical reason why NS does not describe well the more discrete/nonequilibrium case. *As the degree of discreteness/nonequilibrium increases, these non-organized fluxes should receive more attention.*

Because the traditional fluid mechanics theory is based on NS, we are most familiar with NS, and comparing with NS is an important way to recognize and understand the new model, so it is necessary to comb the connection and difference between DBM and NS again: The NS is only a physical model, and itself does not contain any analysis method for the simulation data and the corresponding complex flow field. It considers only the first order effects of the discreteness/TNE described by the Knudsen number, and consequently applies only to the case of quasi-continuum and near-equilibrium. In terms of numerical simulation research, NS is only responsible for determining the governing equations and the necessary constraints before the simulation, not for how to extract helpful information from massive data after the simulation. In short, *the NS is only responsible for the pre-simulation, and not for the post-simulation.* In addition to the description function equivalent to the corresponding level of EHE, DBM also provides a set of complex physical field analysis methods for the massive data and corresponding complex flow field. In terms of numerical simulation research, *the DBM is not only responsible for the pre-simulation, but also for the post-simulation.*

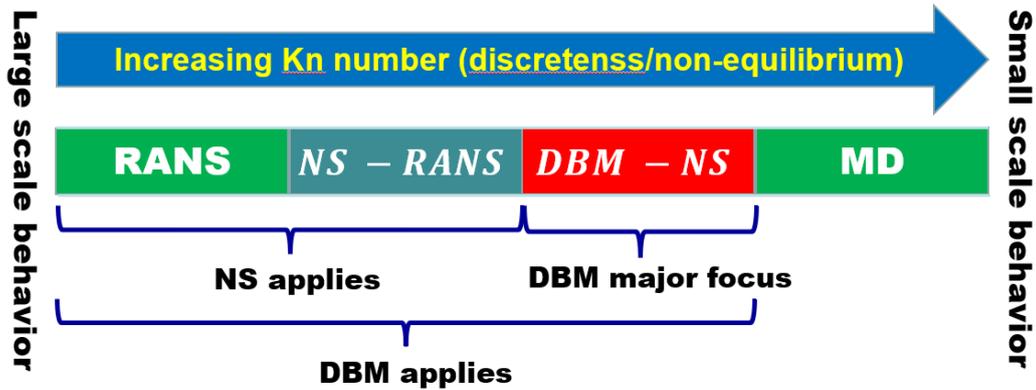

Fig.5: Application ranges of various models

DBM exceeds traditional fluid modeling in terms of application scope and physical function, focusing on the "mesoscale" dilemma of insufficient physical function of macroscopic continuous model and inability of microscopic molecular dynamics simulation due to limited application scale, as shown in Fig. 5. Among them, RANS is short for Reynolds Average Navier-Stokes, which is a coarser-grained physical modeling on the basis of NS and describes the behavior of a large scale within the range of NS. In the figure, "$NS-RANS$" represents the parts poorly described by RANS within the scope of NS application, where "$-$" is a minus sign, indicating that it is subtracted from the scope of application. "$DBM-NS$" indicates the parts of NS that are not well described within the scope of the DBM, which is the major focus of the DBM.

Mesoscale behavior research presents greater challenges in modeling, simulation, and analysis of situations with higher degrees of discreteness/nonequilibrium. The necessity of those most closely related non-conserved moments increases with the degree of discreteness/non-equilibrium. Although mainly aiming at the cases with higher degrees of discreteness/non-equilibrium, nearly every one has to start from and pass through the case of quasi-continuum and near equilibrium. Because everyone needs a congnitive foundation and technical foundation. This is precisely the reason why the field of "mesoscale" behavior research is "slow" to get started and "slow" to progress.

## 4. Examples of application

Because the traditional fluid modeling mainly focuses on the case of quasi-continuum and near equilibrium, the behavior and effect under higher degree of discreteness and nonequilibrium are poorly studies and far from clear. However, it can be thought that these previously poorly understood discrete/non-equilibrium effects must contain a large number of physical functions to be developed.

Entropy production rate is an important physical parameter in many fields related to compression science, such as inertial confinement fusion, aero-engine and explosion damage effect. In inertial confinement fusion, the high entropy generation rate often means that the internal pressure of the target material increases rapidly due to the increase in temperature, resulting in the target material is not easy to be compacted, thus affecting whether the nuclear explosion conditions can be met. In aero-engines and other equipment, a high entropy generation rate often means a large dissipation, so that the system's ability to do mechanical work is low. In the study of explosion damage effect, people have tried to use different physical quantities to characterize the explosion damage effect. Finally, it is found that the entropy increase rate provides a more scientific description method. With DBM, it is convenient to study the main mechanisms causing entropy increase in complex flow processes and their relative importance [13-14].

The TNE behavior features provided by DBM can not only be used to restore the main features of the real distribution function $f$ of the target region [15-16], but also be used for the physical identification of different interfaces and the design of interface tracking technology [17-18]. In the kinetic study of phase separation, the maximum point of TNE intensity $D$ can be used as the first physical criterion for dividing the two phases of Spinodal Decomposition (SD) and Domain Growth (DG) [19]. The maximum point of entropy increase rate $\dot{S}$ can be used as the second physical criterion to divide the two stages of SD and DG [20]. In the kinetic study of hydrodynamic instability, it is possible to observe both material mixing and energy mixing at the same time by examining some TNE intensity, for example, NOEF [21]. The end point of the linear growth phase of NOEF intensity can be used as a physical criterion for the transition from Kelvin-Helmhotz (KH) instability dominance to Rayleigh-Taylor (RT) instability dominance in Rayleigh-Taylor-Kelvin-Helmhotz instability coexistence system [22]. In the kinetic study of droplet collision, shock bubble interaction, and bubble fusion, etc., the TNE behavior characteristics can be used as a physical criterion to distinguish collision types and different stages [23-26]. Interface length $L$ is also one of the dimensions of TNE behavior representation. In wan der Waals fluid RT instability system, the exponential growth stage of $L$ corresponds to the bubble acceleration stage. Both the first maximum point of $dL/dt$ and the first maximum point of the change rate of NOMF induced entropy production rate $\ddot{S}_{\text{NOMF}}$ can be used as the physical criterion for the bubble velocity to enter the progressive stage [27]. In the study of plasma kinetics, TNE characteristics can be used to physically distinguish plasma shock wave, detonation wave and common fluid shock wave. For RM instability systems with interfacial inversion, the magnetic field increases the global average TNE intensity $\bar{D}_T$ before interfacial inversion, but significantly reduces the TNE intensity $\bar{D}_T$ and entropy generation rate after inversion. The asymptotic lower limits of $\bar{D}_T$ and the NOEF entropy production rate $\dot{S}_{NOEF}$ are helpful to determine the effective critical magnetic field in a real system [14], and so on.

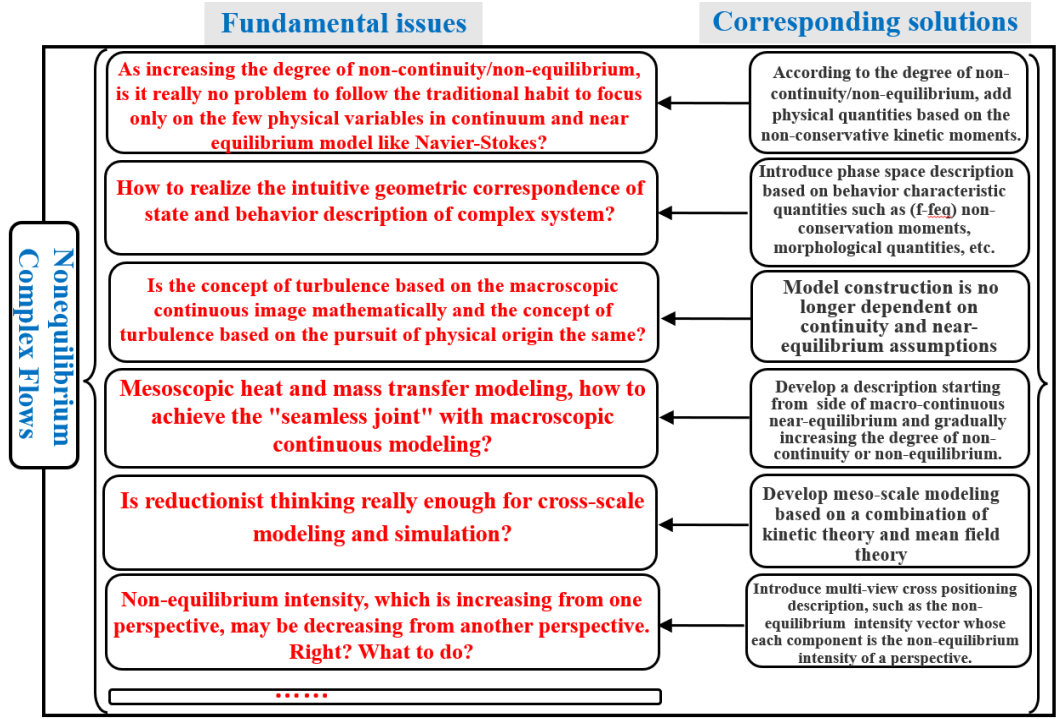

Fig.6 Basic scientific problems resulting in the DBM and corresponding solutions.

## 5. Summary and prospect

Compared with macroscopic behavior, the typical characteristics of mesoscale behavior are as follows: the discrete effect is significant, and the thermodynamic non-equilibrium effect is significant. The description of mesoscale behavior requires more physical quantities. The mesoscale characteristics of DBM include: (i) in terms of scale, it is between micro and macro, and connects micro and macro; (ii) In terms of function: it exceeds NS in the depth and breadth of discrete/non-equilibrium effect description, but is weaker than MD. The basic scientific problems that result in the DBM and corresponding solutions can be summarized in Figure 6.

Mesoscale behavior gives birth to new technology. Research on mesoscale behavior has made gratifying progress, but there is still a long way to go. Fortunately, more and more attention has been paid to the experimental study of non-equilibrium flow. Because the experimental measurement of many effects of non-equilibrium flow is subject to many technical constraints, the progress of experimental research is naturally relatively slow. This is why, in the study of non-equilibrium flow, numerical simulation research is far ahead of experimental research, and this is why numerical simulation research should be first. The practical application of the discrete/non-equilibrium effects is expected to be realized first in numerical experimental research.

**Acknowledgements**：Jiahui Song, et al. helped draw some pictures and proofread the full text. Professor Zhanchun Tu put forward many pertinent opinions and suggestions on the expression. Academician Jinghai Li and Professor Limin Wang gave many enlightening comments.